\documentclass[conference]{IEEEtran}
\IEEEoverridecommandlockouts
\usepackage{cite}
\usepackage{amsmath,amssymb,amsfonts}
\usepackage{algorithmic}
\usepackage{graphicx}
\usepackage{textcomp}
\usepackage{xcolor}
\def\BibTeX{{\rm B\kern-.05em{\sc i\kern-.025em b}\kern-.08em
    T\kern-.1667em\lower.7ex\hbox{E}\kern-.125emX}}
\begin{document}

\title{Towards Productizing AI/ML Models:\\An Industry Perspective from Data Scientists}

\author{
\IEEEauthorblockN{Filippo Lanubile, Fabio Calefato, Luigi Quaranta,}
\IEEEauthorblockA{\textit{University of Bari} \\
Bari, Italy \\
firstname.lastname@uniba.it}
\and
\IEEEauthorblockN{Maddalena Amoruso, Fabio Fumarola, Michele Filannino}
\IEEEauthorblockA{\textit{Prometeia} \\
Milano, Italy \\
firstname.lastname@prometeia.com}
}

\maketitle

\begin{abstract}
The transition from AI/ML models to production-ready AI-based systems is a challenge for both data scientists and software engineers. In this paper, we report the results of a workshop conducted in a consulting company to understand how this transition is perceived by practitioners. Starting from the need for making AI experiments reproducible, the main themes that emerged  are related  to  the use  of  the  Jupyter  Notebook as the primary  prototyping  tool, and the lack of support for software engineering best practices as well as data science specific functionalities. 
\end{abstract}

\begin{IEEEkeywords}
AI/ML models, data science, Jupyter notebooks, data products
\end{IEEEkeywords}

\section{Introduction}

Today, product features based on Machine Learning (ML) are being massively integrated into software systems \cite{amershi_software_2019}. Often, the decision of building AI-based systems comes after data scientists within organizations achieve promising explorative data analysis followed by successful implementation of prototypical models. Once encouraging results are achieved in the lab, the best performing model has to be `productized' --- i.e., turned into an actual `data product,' ready to be used in live services. This transition from AI/ML models to production-ready systems is a complex endeavor, which brings some challenges that still lack consolidated solutions. Many of the obstacles arise from the fact that the main concern of data scientists is the timely delivery of the best performing model \cite{kim_emerging_2016}; they rarely consider building robust and scalable services a priority, and hardly ever deal with the related problems since the earliest stages of a project.

Consequently, an organization that needs to build an AI-based system can take one of two approaches: (1) have the data science team hand-off the model to a software engineering team, which take over the task of integrating it into production; (2) let the data science team handle the whole model lifecycle, including its integration. While the former approach entails unavoidable inefficiencies, due to the necessary porting and reworks, as well as the typical shortcomings in the communication between data scientists and software engineers \cite{sculley_hidden_2015, marin_data_2019}, the latter is easily limited by the tools and practices of data scientists, which typically lack support for production-level development. One example is the computational notebook -- and in particular, Jupyter Notebook \cite{perkel2018jupyter, perez_project_2015}: this is the tool of choice for many data scientists to perform data exploration and model building, as its interactivity is particularly convenient for such tasks, but it offers poor native support even for basic collaborative software development practices such as code versioning, modularization, and testing.

One of the main challenges that data scientists face when they translate their prototypes into products is the reproducibility of the analysis and model training process \cite{lwakatare_taxonomy_2019}. Nonetheless, ensuring the reproducibility of AI/ML code is nontrivial: on the one hand, some algorithms have an intrinsically non-deterministic nature; on the other hand, the complex pipelines set up by data scientists to build models often depend on untracked environmental dependencies, intricate and scarcely-portable system configurations, and rely on tools that can hardly handle reproducibility by themselves, as in the case of computational notebooks \cite{pimentel_large-scale_2019}.
Reproducibility should receive special attention because it is crucial not only for the reliability of AI-based systems but also for their maintainability \cite{sculley_hidden_2015}.

To gain a better understanding of this challenge and figure out how it is perceived by practitioners, we organized a workshop at an Italian consulting company, 
involving data scientists with different experience levels and educational backgrounds. In this paper, we report the main themes that emerged from the discussion: the benefits and pitfalls of computational notebooks, the desire for notebook features and tools supporting the reproducibility of AI/ML experiments, and the need for shared best practices.

The remainder of this paper is organized as follows: in Section~\ref{sec:design&execution}, we describe the workshop organization; in Section~\ref{sec:analysis&discussion}, we report and comment on the main themes that emerged from the workshop; in Section~\ref{sec:conclusion}, we conclude the paper with a hint to future work.

\section{Workshop Design and Execution}
\label{sec:design&execution}


\subsection{Business and Social Context}

Prometeia is an Italian provider of consulting services, software solutions, and economic research focused on risk, wealth, and asset management. The company has more than 300 clients in over 20 countries around the world, mainly banks, insurance companies, and institutional investors. 

We organized the workshop in the Milan branch, which is where the data science team is mainly based. Eighteen data scientists attended the event (3 women and 15 men); four participants attended remotely, from the headquarter of the company in Bologna.
Workshop attendees have a varied educational background: some of them have a M.Sc. in Computer Science, while others have studied Physics, Mathematics, and Statistics. 

\subsection{Workshop Set Up}

We held the workshop in the Fall 2019. The company reserved a comfortable room for the event, where we arranged chairs around a large table, leaving room for the projector screen on one side. 
Before the beginning of the meeting, we checked that the connection with the remote attenders worked. We set up a camera to let them see the room alongside the presentation slides.

One industry author was in charge of facilitating the discussion, 
while one academic author was in charge of the presentation and another took part as the note taker. Together with textual notes, we collected audio and video recording of the entire event.

\subsection{Workshop Execution}

The workshop lasted 2 hours and a half.
During the first hour, after welcoming the audience and giving a brief synopsis of the session, the presenter outlined his review of the currently available software solutions for AI experiments, focusing on the most relevant examples. 

During the presentation, the speaker involved the audience multiple times, collecting information about their acquaintance with the presented tools through interactive polls. As tools were illustrated, the audience showed considerable involvement by asking questions on specific aspects of their usage, mainly expressing concerns about data management issues (privacy, data lock-in, etc.). They explained that they worked in a mission critical context in which data confidentiality is almost always a priority.

The presenter concluded his talk with the following three questions, aimed at triggering the discussion in the subsequent brainstorming session:

\begin{itemize}
	\item ``Which are your favorite tools and why?''
	\item ``Are there any gaps between tool features and your state of practice/needs?''
	\item ``Should we use notebooks only for reporting early results and prototyping?''
\end{itemize}

In the first part of the brainstorming session, the moderator encouraged the participants to freely take part in the discussion, either by answering one or more questions regarding the presentation or by sharing their views on the topic. After hearing most of the attendees, to ensure that we listened to all the voices in the audience, the moderator proposed that everyone in the group introduced themselves -- providing information on their educational background and role -- and freely expressed some thoughts on the topic.

Overall, as detailed in the following section, not only we were able to collect a number of interesting comments on the theme of reproducibility in AI experiments, but also we gathered precious insights about (1) the role of Jupyter Notebook in the typical workflow of the involved practitioners, (2) the desired product features and tools that would solve some of the most pressing challenges currently faced by the company and (3) the need for validated and shared best practices to be adopted by the data science team.

\section{Analysis and Discussion}
\label{sec:analysis&discussion}


\subsection{Method}

Shortly after the workshop, the moderator and the note-taker conducted a quick debriefing session, to ensure that all relevant aspects of the discussion had been noted down and add any further comments or impressions. Later on, the note-taker went through the recorded material, integrating field notes where necessary and transcribing the most informative quotes from the brainstorming.

Eventually, we reviewed all the notes and performed a thematic analysis to extract and group the most relevant topics from the discussion. 

\subsection{Themes emerged from the workshop}


\subsubsection{The Role of Computational Notebooks}

The question on the role of Jupyter notebooks was the one that immediately catalyzed the attention of the participants, thus showing us that the use of notebooks was already a hot topic in the team. One of the attendees started by saying that a large part of the work they currently do at Prometeia has an intrinsically exploratory nature. Most of the time, prototyping plays a major role and Jupyter Notebook ends up being a particularly convenient tool to use.

\textbf{It is out of the question to stop using computational notebooks.} Another participant firmly argued that none of the tools from the presentation could ever be adopted by the company if that meant to dismiss computational notebooks. Indeed, once a data scientist develops a good method for writing notebooks, they compose them in such a way that helps their future self and colleagues to fully reconstruct the history of a successful experiment:

\begin{quote}
    ``This `storytelling' capability of notebooks is even more appealing and useful than the certainty that experiments will be reproducible. Not to mention that the interactivity of Jupyter Notebook is simply too much valuable for rapid and effective prototyping... nobody would give it up!''
\end{quote}

\textbf{An effective development environment should combine computational notebooks and structured codebases.} Four data scientists agreed that computational notebooks alone cannot replace traditional, structured codebases; instead, they should coexist and be used as complementary tools. If notebooks are a good entry-point for learning data science and an optimal prototyping tool, they simply do not scale to production. One of the participants affirmed:

\begin{quote}
	``A solution might be to use the best of both worlds: the interactivity of computational notebooks and the solidity of structured codebases. For example, after experimentations have been carried out using a notebook and the desired outcome for a particular phase has been achieved, a developer should transfer the code from the notebook itself to a structured repository. This way, the portions of code that are actually useful to productize the work gradually settle down in a well structured and testable environment, ready to evolve over time.''
\end{quote}

This idea that `a developer should transfer the code from the notebook to a structured codebase' was taken up by another colleague, who added:

\begin{quote}
	``This kind of work might be effectively performed by a pair of data scientists (or a data scientist and a software engineer): the first should be concerned with data exploration using computational notebooks, while the second should do the job of layering validated code in a production codebase.''
\end{quote}

\textbf{Notebooks should be dropped at a certain point in the workflow.} Some of the participants were more categorical in saying that notebooks should be dropped at a certain point in the workflow, as some of their intrinsic characteristics (e.g., non-linear execution of code) can easily become a trap while the code keeps growing. One even affirmed that the goal should be to reduce the use of notebooks as much as possible and to dismiss them since the earliest phases of a project: 

\begin{quote}
	``Serious production code cannot but be developed in its own classical environment, which is the result of almost 40 years of experience in software development and engineering.''
\end{quote}


As can be noted, the range of opinions on the role of computational notebooks is quite varied; nonetheless, no one puts their importance into question. In contrast, disagreement among the practitioners lies in the identification of the right step in the workflow where notebooks should give way to standard code -- i.e., where code quality and reproducibility should be assigned a higher priority than exploration ease and fast prototyping.

\subsubsection{Dream Notebook Features and Tools} During the brainstorming session, many of the workshop participants expressed the desire for specific notebook features and further tools supporting their AI/ML workflow.

\textbf{Testing support during data exploration.} Some data scientists underlined the lack of support for code testing in the early explorative phase of a project. To enable analysis reproducibility, one should be at least able to test their code for correctness. Notebooks themselves do not offer native support for unit testing and assertions, and yet they should:

\begin{quote}
	``Notebooks ought to evolve and become more like IDEs in this sense, or at least to represent a better springboard for IDEs.''
\end{quote}

As a result, at the beginning of data science projects, testing is almost regularly skipped and becomes an actual concern only later, when software engineers consolidate code into standard Python packages.

One of the participants pointed out that notebooks should be automatically checked before being pushed to a shared repository, e.g. by using a CI automation server like Travis CI and Jenkins. To be uploaded, they must be at least runnable without errors.

\textbf{Experiment versioning support.} Being inspired by Donald Knuth's literate programming paradigm, notebooks support `computational narratives', that is, the narrative description of analytical processes. They provide a lightweight form of documentation that leaves a trail on decision rationale, interpretation of results, etc. However, most of the time, one single successful computation results from a long series of explorations and failures. A couple of attendees observed that notebooks should push the support for computational narratives even further by storing an explicit history of the whole trial-and-error process leading to the desired solution. Each exploration/modeling attempt should be saved separately, with comments describing it. Branching should be supported and each saved step in the explorative process should be retrievable through a visual interface. Obviously, this could be achieved by employing traditional code versioning tools; however, such tools are not natively supported in notebooks. Jupyter has its own checkpoint system to provide lightweight versioning of notebooks. However, neither branching nor version commenting are supported. On the other hand, versioning a notebook via git is possible, but diffs are difficult to read, as notebooks are specified in the \texttt{.json} format. One has to resort to external tools, like Project Jupyter's \texttt{nbdime}\footnote{https://github.com/jupyter/nbdime}, to visualize the differences with more ease. Nevertheless, none in the audience mentioned the need for a notebook-specific diffing tool.

Another couple of participants pointed out the importance of data versioning. At the time of the workshop, they used carefully crafted folder structures and naming conventions to handle different versions of the same data. Not only it was time-consuming but also error-prone; additionally, they had no way to easily spot the differences between two versions of the same large dataset. Data versioning tools should allow straightforward visualization of data diffs, at least in a summarized form (e.g., `2 lines added; 30 lines removed'), and they should offer typical git-like features, as the possibility to comment commits and to push single data versions to remote storage services. Here we notice that, even if not integrated with Jupyter Notebook, DVC\footnote{https://dvc.org} (a git-inspired tool reviewed during the presentation) fully covers the desired set of functionalities, although it might have a steep learning curve, especially for data scientists that have had no previous experience with git.

\textbf{Reproducibility aids in notebooks.} One of the main threats to reproducibility in computational notebooks is the arbitrary execution order of code cells. It might be difficult to reconstruct the right execution order, especially if some of the cells have been run repeatedly. Messy executions often lead to hidden states, i.e., states of the Python interpreter that cannot be inferred and restored from the evidence left in the notebook code. One of the participants suggested that notebooks should record execution sequences and cell activations to provide automatic (or semi-automatic) inference of the right execution order.

Another typical area of notebook reproducibility failure is dependency management \cite{pimentel_large-scale_2019}. One of the attendees witnessed that, although Python and R are portable programming languages, porting code from one machine to another is not always straightforward, the main reason being project dependencies: for instance, different operating systems might support different versions of the same Python library. To complicate matters further, it is not rare that the client defines the specific library versions to be used in a project. To ensure compliance with the requirements and reproducibility, dependencies should be always explicitly declared in dedicated requirements files.

It appears evident that, at least for this company, improving the transition from prototypes to products and the reproducibility of AI pipelines is closely related to the use and capabilities of the primary prototyping tool, i.e., the computational notebook. Jupyter Notebook is supposed to evolve by mitigating its peculiar threats to reproducibility (e.g., untracked non-linear executions and the consequent formation of hidden states); moreover, native support is highly desired for traditional software engineering best practices (like code testing and versioning), and data science specific functionalities (e.g., data versioning).

\subsubsection{The Need for Shared Best Practices}

One of the participants, a data scientist with a background in computer science, introduced a further theme in the discussion:

\begin{quote}
	``How much of the reproducibility issues are due to the inadequacy of tools, and how much of them are to be imputed to developers’ bad practices?''
\end{quote}

A senior data scientist replied that the definition of a validated set of best practices should be a priority for the team. Indeed, an informed choice of specific tools to improve the workflow can be done only after the identification of optimal shared behaviors; it should naturally happen as a consequence of it.

One of the remote attendees added:

\begin{quote}
	People must learn to carefully follow best practices. I strongly believe in the role of education and personal sensitivity to such kind of issues. Often, the required awareness of these topics is low in people coming from non-CS backgrounds.
\end{quote}

These words made explicit another common problem: the challenging collaboration between data scientists and software engineers. Often, these roles have different educational backgrounds and diverse levels of awareness of software engineering issues. A possible approach to face this challenge is suggested by the other remote attendee, who concluded:

\begin{quote}
	``Knowledge and expertise sharing between the actual data scientists -- purely devoted to modeling -- and machine learning engineers -- concerned with models in production -- is essential. A single tool to be shared by both these figures, encouraging best practices, would be ideal.''
\end{quote}

We advise the importance of shared training sessions aimed at closing the gaps that currently limit the collaboration between these two complementary roles. Furthermore, a set of shared best practices should be carefully developed at a company level and refined over time; the strategy might involve field studies to validate the most promising habits and regular team retrospectives to collectively assess the results.

\section{Conclusion}
\label{sec:conclusion}

In this paper, we reported the insights gathered from a workshop on productizing AI/ML models, held at a consulting company.  
We conducted the workshop at a point in time where the Data Science team was in its infancy. Today's team has almost doubled the number of its members for several roles such as data scientists, big data engineers and business translators. The workshop has contributed to trigger an evolution in terms of tools and best practices that has led to the creation and adoption of a system for ingesting, analyzing and reporting on data: the Prometeia Modeling Platform.
As future work, we envisage the need for a field study to determine the optimal lifecycle for a computational notebook and develop a cohesive view of the role and objective limitations of this tool in the context of a cross-functional team including both data scientists and software engineers.

\ifCLASSOPTIONcompsoc
  \section*{Acknowledgments}
\else
  \section*{Acknowledgment}
\fi
We would like to thank all data scientists that participated in the workshop and provided insightful comments. 

\bibliographystyle{IEEEtran}
\bibliography{IEEEabrv, references}

\end{document}